\documentclass[a4paper,fleqn,usenatbib,10pt]{mnras}

\usepackage{mathptmx}

\usepackage[T1]{fontenc}
\usepackage{ae,aecompl}

\usepackage{amssymb,amsmath,color,verbatim,graphicx,url,algorithm2e}

\title[\textsc{The-wiZZ}]{\textsc{The-wiZZ}: Clustering redshift estimation for everyone}

\author[C. B. Morrison et al.]{
C. B. Morrison,$^{1,2}$\thanks{E-mail: morrison.chrisb@gmail.com}
H. Hildebrandt,$^{1}$
S. J. Schmidt,$^{3}$
\newauthor
I. K. Baldry,$^{4}$
M. Bilicki,$^{5}$
A. Choi,$^{6,7}$
T. Erben,$^{1}$
and P. Schneider$^{1}$
\\
$^{1}$Argelander-Institut f\"ur Astronomie, Auf dem H\"ugel 71, 53121 Bonn, Germany\\
$^{2}$Department of Astronomy, University of Washington, Box 351580, Seattle, WA 98195, USA\\
$^{3}$Department of Physics, University of California, Davis, One Shields Ave., Davis, CA 95616, USA\\
$^{4}$Astrophysics Research Institute, Liverpool John Moores University, IC2, Liverpool Science Park, 146 Brownlow Hill, \\Liverpool, L3 5RF, UK\\
$^{5}$Leiden Observatory, Leiden University, P.O. Box 9513 NL-2300 RA Leiden, The Netherlands\\
$^{6}$Scottish Universities Physics Alliance, Institute for Astronomy, University of Edinburgh, Royal Observatory, Blackford Hill, \\Edinburgh, EH9 3HJ, UK\\
$^{7}$Center for Cosmology and Astro Particle Physics, The Ohio State University, 191 W. Woodruff Avenue, \\Columbus, OH 43210, USA
}

\date{Accepted XXX. Received YYY; in original form ZZZ}

\pubyear{2017}

\begin{document}
\label{firstpage}
\pagerange{\pageref{firstpage}--\pageref{lastpage}}
\maketitle

\begin{abstract}
We present \textsc{The-wiZZ}, an open source and user-friendly software for estimating the redshift distributions 
of photometric galaxies with unknown redshifts by spatially cross-correlating them against a reference sample 
with known redshifts. The main benefit of \textsc{The-wiZZ} is in separating the angular pair finding and 
correlation estimation from the computation of the output clustering redshifts allowing anyone to create a 
clustering redshift for their sample without the intervention of an ``expert''. It allows the end user of a given 
survey to select any  sub-sample of photometric galaxies with unknown redshifts, match this sample's catalog 
indices into a value-added data file, and produce a clustering redshift estimation for this sample in a fraction of 
the time it would take to run all the angular correlations needed to produce a clustering redshift. We show results 
with this software using photometric data from the Kilo-Degree Survey (KiDS) and spectroscopic redshifts from 
the Galaxy and Mass Assembly (GAMA) survey and the Sloan Digital Sky Survey (SDSS). The results we 
present for KiDS are consistent with the redshift distributions used in a recent cosmic shear analysis from the 
survey. We also present results using a hybrid machine learning-clustering redshift analysis that enables the 
estimation of clustering redshifts for individual galaxies. \textsc{The-wiZZ} can be downloaded at \url{http://
github.com/morriscb/The-wiZZ/}.
\end{abstract}



\section{Introduction}

Current and future photometric galaxy surveys are designed to measure the properties and evolution of 
galaxies as well as constrain cosmological parameters and the properties of the Universe. In order to enable 
this, accurate and unbiased estimates of galaxy redshifts are required to extract the maximum amount of 
information. Until recently, redshift information in photometric surveys was only gained through spectroscopic 
followup or photometric redshifts (photo-$z$s) from multi-band photometry. Many techniques exist for deriving 
photo-$z$s \citep[see][for a partial review]{hildebrandt10}, however all these techniques rely on a calibration set 
of spectroscopic redshifts that is representative of the survey galaxy population. Such a sample of spectra is 
only possible for the shallowest surveys and still requires a significant amount of telescope time. For future 
deep, large-area surveys such as The Large Synoptic Survey Telescope\footnote{\url{http://www.lsst.org/}} 
(LSST), a sample of representative spectra will be even more difficult. Such challenges are presented in 
\citet{newman15}.

An alternative and complementary method to photo-$z$s is that of clustering redshift estimation
(clustering-$z$s). Clustering redshifts make use of the fact that galaxies with unknown redshifts reside in the 
same structures as galaxies that have known redshifts. Thus, spatial cross-correlations can be used to estimate 
the redshift distribution of the sample with unknown redshifts. The basic method bins the sample with known 
redshifts in $z$ and then spatially cross-correlates each of these bins against the unknown sample. The 
amplitude of the resultant correlation can then be used to estimate the amount of redshift overlap and thus the 
redshift distribution of the sample with unknown redshifts. One of the first suggestions of such a method can be 
seen in \citet{schneider06} with the formalism for this method written out in \citet{newman08} and later 
generalized in \citet{schmidt13} and \citet{menard13} with quadratic estimators laid out in \citet{mcquinn13} and 
\citet{johnson16}. The method has some drawbacks from sensitivity to galaxy bias both from the reference 
sample with known redshifts and the sample with unknown redshifts which can affect clustering-$z$s. However, 
suggestions to mitigate this bias exist in the literature \citep{newman08, menard13, schmidt13}.

Cross-correlation techniques are beginning to be applied to real data \citep{rahman15, choi15, rahman16a, 
rahman16b, scottez16, hildebrandt16, johnson16} with an eye towards future surveys. A failing of this method, 
however, is that the current implementations of clustering redshifts are not as easy to use as their photo-$z$ 
counterparts and nominally require spatial correlations to be run and re-run for each galaxy sub-sample of 
interest. This is a time-consuming process and could limit clustering redshift's adoption by the larger community. 
Suggestions exist such as producing clustering-$z$s in color-color space cells \citep{rahman16a, scottez16} but 
this will have limitations for some samples and precludes the ability to weight galaxies in the clustering redshift 
estimation in the same way as in a given  analysis or utilize additional information after the correlations are run 
for each cell. A more flexible method that separates the spatial correlation computation from the act of creating 
clustering redshifts would be ideal.

In this article we present \textsc{The-wiZZ}\footnote{Available at: \url{http://github.com/morriscb/The-wiZZ/}}, a
method for estimating redshift distributions from clustering designed for ease of use by survey end users. 
\textsc{The-wiZZ} separates the difficult step of finding close angular pairs from the act of creating a clustering 
redshift estimate. In this way the correlations between close pairs can be run once by the survey data pipeline 
and then an end user can create a clustering redshift estimate for their unique sub-sample of galaxies in a 
matter of, in some cases, seconds. \textsc{The-wiZZ} can add to the legacy of galaxy surveys by producing a 
stable data product that can continue to be used by the astronomy community without a large amount of 
specialized software, much like how photo-$z$s are used today. \textsc{The-wiZZ} can of course also be used 
by individuals with any data overlapping a spectroscopic sample allowing them to produce clustering-$z$s 
quickly and easily.

This document is laid out as follows. In Section \ref{sec:method} we give an overview of the method and
software including showing how it can be used in the context of a galaxy survey. Section \ref{sec:data} explains 
the data products we use to test \textsc{The-wiZZ}. In Section \ref{sec:recoveries} we show the resultant 
clustering redshift estimates and present a novel method of color-redshift mapping made possible by the speed 
of \textsc{The-wiZZ}. Section \ref{sec:discussion} discusses these redshift estimates and \textsc{The-wiZZ} in 
the context of current surveys. Finally in Section \ref{sec:conclusions} we present our conclusions with an eye 
toward future surveys such as LSST, Euclid\footnote{\url{http://sci.esa.int/euclid/}}, and The Wide Field Infrared 
Survey Telescope\footnote{\url{http://wfirst.gsfc.nasa.gov/}} (WFIRST). Throughout this analysis we use the 
WMAP5 \citep{komatsu09} cosmology for consistency between the code we use for spatial pair finding, 
\textsc{STOMP}\footnote{Available at: \url{http://github.com/ryanscranton/astro-stomp/}}, and \textsc{The-wiZZ}. 
The choice of cosmology will, however, have little effect on the resultant clustering-$z$s \citep{newman08, 
matthews10}. 


\section{Method Overview}\label{sec:method}

The methodology of \textsc{The-wiZZ} is to separate the computationally intensive step of pair finding and 
angular correlation estimation from the creation of a clustering redshift estimate for a given galaxy sample of 
unknown redshift, allowing for fast computation and re-computation of the output clustering redshift estimate. 
We do this by pre-computing and storing all pairs between a galaxy sample with known redshifts (hereafter 
known as the reference sample) and catalog of galaxies with unknown redshifts (hereafter, the unknown 
sample) within a fixed physical radius around the reference galaxy. This is similar in concept to fast correlation 
codes pre-computing data structures for quick pair finding/correlation estimation. End users can then simply 
select their desired sub-sample from the unknown sample catalog and match the catalog indices of their sample 
into the data file containing the pairs using the provided software. \textsc{The-wiZZ} then takes care of all the 
book keeping and produces a properly normalized estimate of the sub-sample's over-density as a function of 
redshift which can then be converted into a clustering redshift estimate or estimated probability density function 
(PDF).

\textsc{The-wiZZ} is thus extremely powerful for use within survey collaborations and as a legacy, value-added 
catalog data product for users of the survey's data in the future. The software is designed to make creating 
clustering redshifts for any unknown sample nearly as easy as selecting in photo-$z$. This is especially 
powerful in the context of survey collaborations as each working group will likely have their own selections and 
weighting scheme for optimal signal to noise within the context of the science they are interested in. Without
\textsc{The-wiZZ} this would require computation of the angular-correlations and clustering-$z$s for each 
unknown sample in question, and if the samples a working group was using ever changed the clustering-$z$s 
would have to be computed all over again. \textsc{The-wiZZ} circumvents this problem by effectively computing 
the correlations for all galaxies in the unknown sample against the reference sample simultaneously, collapsing 
these measurements into a clustering redshift only when called with a user specified sub-sample of galaxies.  
Additional data in newly observed areas can be easily appended in this data structure without having to rerun the 
full sample. The only time the pair finding portion of \textsc{The-wiZZ} would have to be re-run is if the 
photometric detection catalog of the survey were to fundamentally change (e.g. new detection algorithm, new 
thresholding, increased survey depth). \textsc{The-wiZZ} makes use of mostly widely available and 
well supported packages including the \textsc{Python} based astronomy library \textsc{astropy}
\footnote{Available at: \url{http://www.astropy.org}} \citep{astropy} making it even easier for end users to get set 
up and started.

\begin{figure}
    \includegraphics[width=0.5\textwidth]{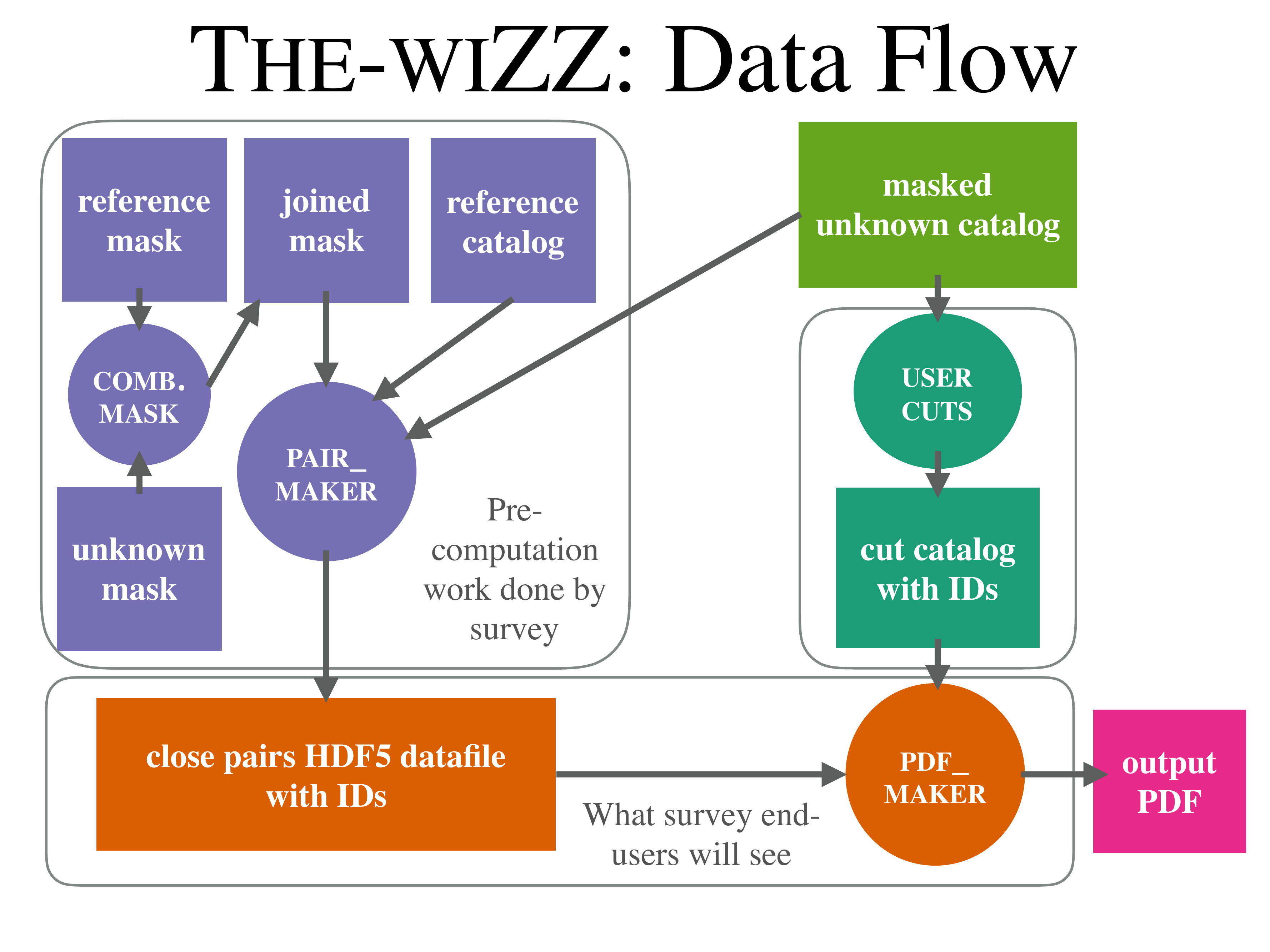}
    \caption{\label{fig:wizz_flow}
      Flow cart of the inputs and output of \text{The-wiZZ}. In the upper left we have the work done by an 
      individual survey in spatially masking the catalog, running \textsc{pair\_maker}, and creating 
      \text{The-wiZZ}'s output \text{HDF5} data file. The upper right shows a user selecting a sample from the 
      masked catalog for their own work. The lower portion is the end user matching their specific sample into the 
      data file using \textsc{pdf\_maker} and producing a resultant clustering redshift estimate for their sample 
      without having to run any cross-correlations.}
\end{figure}

Fig. \ref{fig:wizz_flow} shows the data flow through \textsc{The-wiZZ} with pre-computation of the correlations 
on the left and the user selected catalog on the right. The lower panel shows the part of \textsc{The-wiZZ} that 
takes the precomputed angular correlations and matches them with a catalog of their specific unknown sample 
to produce an output clustering redshift distribution. This is the part of \textsc{The-wiZZ} that the majority of 
users will see.

The remaining portion of this section goes into depth about the internals of both the pre-computation step 
\textsc{pair\_maker} and the final clustering redshift estimation creation step \textsc{pdf\_maker}.


\subsection{\textsc{pair\_maker}}

The left hand side of the data flow diagram shown in Figure \ref{fig:wizz_flow} shows the input and outputs of 
the \textsc{The-wiZZ} program called \textsc{pair\_maker}. The program does the majority of the calculations 
involved in masking catalogs, finding pairs, and generating random points. The method we utilize is described in 
detail in \citet{schmidt13} and \citet{menard13}. We will describe some of our modifications and generalizations 
to that method in this subsection. The \textsc{pair\_maker} software utilizes the STOMP spherical pixelization 
library for masking and pair finding. Further information on STOMP can be found in \citet{scranton02} and 
\citet{scranton05}.

Our first step in using \textsc{pair\_maker} is creating a combined mask of the area covered by both the 
unknown and reference samples. STOMP stores these resultant Maps in a hierarchical pixel format that allows 
for storage of pixels at different resolutions for optimal file size and quick spatial searches. This masking Map 
contains the observed area of the survey with bright stars, bad pixels, etc. masked out. Once this Map is 
created we can load both our reference and unknown samples with their indices into \textsc{The-wiZZ}. 
\textsc{pair\_maker} then creates a searchable quadtree on the STOMP hierarchical pixels using the unknown 
sample and also generates random points on the mask if requested and stores them in a similar quadtree.

\begin{algorithm}
  \KwData{reference catalog, unknown catalog, $R_{\rm min}$, $R_{\rm max}$, randoms, data file}
  \KwResult{Stored, sorted unknown ids and number of randoms, per reference object with meta-data}
  \For {object in reference catalog} {
    find object spatial region\;
    find pair ids and distances within $R_{\rm min} - R_{\rm max}$\;
    sort ids and distances by id\;
    find $n$ randoms within $R_{\rm min} - R_{\rm max}$\;
    store sorted ids and distances in data file\;
    store $n$ randoms in data file\;
    store object redshift in data file\;
    store object spatial region in data file\;
  }
  \caption{\label{algo:pair_finder}
  Pseudo-code describing the main loop of \textsc{pair\_maker}.}
\end{algorithm}

Unlike \citet{matthews10}, \citet{choi15}, or \citet{johnson16} in which they measure and fit the full angular 
correlation function in bins of redshift for the reference sample we only measure the correlation amplitude in a 
single bin in projected radius. STOMP allows for easy conversion from a fixed radial bin in physical radius to an 
angular bin on the sky given the redshift of the reference object and an assumed cosmology. We use this binning 
for all measurements with \textsc{The-wiZZ}. STOMP then finds all the pixels at a fixed resolution that cover this 
annulus and uses these pixels to search the unknown quadtree. \textsc{The-wiZZ} then stores the unique index 
of each unknown object as well as the inverse distance from the reference object to the pixel center. One of the 
key aspects of the method presented in \citet{schmidt13} and \citet{menard13} is the signal-matched filtering and 
weighting of the galaxy pairs by the inverse projected physical distance to the reference object. \citet{newman08} 
and \citet{matthews10} show that clustering redshift measurements depend only weakly on the assumed 
cosmology so the clustering-$z$s \textsc{The-wiZZ} produces can be assumed to be general. This pair 
information is then stored for each unmasked reference object. If requested, the software repeats the process for 
the random sample but only stores the total number of randoms found and the sum of their inverse distances 
rather than storing the information for individual galaxies. These steps are sketched out in pseudo-code in 
Algorithm \ref{algo:pair_finder}.

\textsc{pair\_maker} can then repeat this process for a large number of requested radial bins using the same 
masked data set and randoms. For this analysis we use a binning of $R=100-1000 \rm{kpc}$. We also 
computed bins similar to those of \citet{schmidt13} with physical radius bins of
$R=3-30;30-300;300-3000\,\rm{kpc}$. We combine some of these bins as $R=3-300;30-3000\,\rm{kpc}$. Note 
that bins of abutting radii are not completely independent from one another. Given the coarse pixelization of 
STOMP in finding pairs the bins are likely to overlap slightly. (i.e. the $R=3-300\,\rm{kpc}$ bin is not simply the 
addition of its two child bins). Computing multiple scales allows the user to find the correct compromise 
between reduced sensitivity to non-linear galaxy bias (large scales) and signal to noise (smaller scales). The 
choice will likely depend on the sample used. Combining multiple scales could also be used as in the 
quadratic-estimators of \citet{mcquinn13} and \citet{johnson16}.

As an aside it should be stated that any functional form of weighting by distance is possible with
\textsc{The-wiZZ}. Weighting by the inverse distance is conceptually simple and close to the roughly expected 
power law scaling of the correlation galaxy function of $\gamma = 1.8$. The software allows for simple 
modifications of this weighting scheme and can be extended to any weighting as a function of projected physical 
distance. This weight function could be modified, for instance, to a similar weighting of \citet{mcquinn13} or 
\citet{johnson16} that attempts to optimally weight for number of galaxies and mitigation of non-linear scales. 
We leave it up to the user to decide what is best for their analysis with the default behavior being inverse 
distance.

The output from the pair finding processes is stored in a custom data structure in \textsc{HDF5} format for later 
use with \textsc{pdf\_maker}. Intricate knowledge of this format or how it is used is not required to utilize 
\textsc{The-wiZZ}. The unique indices of each unknown object as well as their inverse distance from the 
reference object are stored in sorted arrays for each reference object, for each scale considered. Several other 
data products are stored per reference object such as its redshift and the number of randoms around the object. 
We attempt to reduce the final file size through lossless data compression. In the end the final size of the data 
files depends on the scales requested, number of reference objects, and the density of the unknown sample. As 
an example, the data file created for the analysis we present in Section \ref{sec:recoveries} is roughly equal in 
size compared to the input unknown catalog masked to the area covered by the reference sample. Currently 
this size comes from using very simple and straight forward techniques and data structures to output the 
resultant pairs. This ratio of input catalog to output is likely to improve as better and more efficient techniques 
are applied to the storage of the data.


\subsubsection{Notes on STOMP regions}

STOMP contains powerful internal methods for creating regions on the sky for spatial bootstrapping and 
jackknifing. Such regionation can be difficult given a complex survey mask and the requirement that regions be 
equal area and regular in shape. These regions are extremely useful for mitigating the effects of observing 
strategy and the density systematics that come with them. STOMP allows for the creation of regions that are 
roughly square and equal area. STOMP regions are what \textsc{The-wiZZ} uses to compute spatial bootstrap 
errors on the clustering-$z$s and are thus extremely important. \textsc{The-wiZZ} also uses said regions to 
significantly speed up the pair matching in \textsc{pdf\_maker}. One should specify regions that are a 
compromise between observational errors and the scales desired. For instance, it may not be possible to run a 
scale that is larger than the size of individual pointings in a multi-epoch survey. The user of the software is 
encouraged to experiment with this variable for their own survey.


\subsection{\textsc{pdf\_maker}}

\textsc{pdf\_maker} is the part of \textsc{The-wiZZ} that the large majority of users will interact with. It is the 
portion of the codebase that takes the resultant \textsc{HDF5} data file created from \textsc{pair\_maker} and 
combines it with the user's sub-sample and returns the clustering redshift estimation. The right hand side of 
Figure \ref{fig:wizz_flow} shows the work a user of \textsc{pdf\_maker} will perform in preparing to utilize 
\textsc{The-wiZZ} for creating clustering redshifts. The user selects a sub-sample of galaxies from the same 
catalog that was masked and used in \textsc{pair\_maker}. The user then invokes \textsc{pdf\_maker} with this 
sub-sample and the \textsc{HDF5} data file output from \textsc{pair\_maker} to create a clustering redshift 
estimate for their specific sub-sample as in the lower portion of Figure \ref{fig:wizz_flow}. At the run time of 
\textsc{pdf\_maker} the user requests one of the scales stored in the \textsc{HDF5}, \textsc{pair\_maker} data 
file and a redshift binning. \textsc{pdf\_maker} then computes the natural estimator of over-density 
\citep{davis83}
\begin{equation}
\delta(z_i) = \frac{D_{\rm r}D_{\rm u}(z_i)}{D_{\rm r}R(z_i)} - 1
\end{equation}
where $D_{\rm r}D_{\rm u}(z_i)$ are the pairs between the reference (${\rm r}$) and unknown (${\rm u}$) 
sample in redshift bin $z_i$. $D_{\rm r}R(z_i)$ are the pairs between the reference sample and random 
positions draw from the same mask as the unknown sample. During this calculation, the number of randoms are 
properly scaled to the requested sub-sample and any weights requested for the unknown sample are applied 
(e.g. shape weights, detection efficiency, photometric redshift posterior probabilities).

\begin{algorithm}
  \KwData{sample catalog, data file}
  \KwResult{over-density around each reference object}
  \For {ref-obj in data file} {
    load stored unkn-ids around ref-object\;
    rescale $n$ randoms around ref-object to match sample\;
    set $n$ unkn-objs around ref-obj to zero\;
    \For {unkn-id in sample catalog} {
      binary search for unkn-id in unkn-ids around ref\;
      \If{unkn-ids contains unkn-id}{
        add 1 to $n$ unkn-objs\;
      }
    }
    store unkn-objs divided and scaled n randoms\;
  }
  \caption{\label{algo:pdf_maker}
  Pseudo-code describing the main loop of \textsc{pdf\_maker}.}
\end{algorithm}

\textsc{The-wiZZ} minimizes the amount of time spent matching pairs from the user specified sub-sample by 
sorting the IDs of the sub-sample and matching them into the, already sorted, IDs stored around each reference 
object using a binary search tree. Algorithm \ref{algo:pdf_maker} shows in pseudo-code the steps
\textsc{pdf\_maker} performs. The software also makes use of two methods of spatially locating the pairs for 
matching. First, \textsc{The-wiZZ} takes advantage of the fact that many source detection programs return IDs 
that are partially sorted spatially. For instance, SExtractor \citep{sextractor} returns IDs that are ordered in 
increasing $y$-axis position and then increasing $x$-axis position such that a sub-selection of increasing, 
ordered IDs  will be localized between a $x$-min and $x$-max and thus localized spatially. The software 
recognizing this results in a speed up of the analysis by a moderate amount, but spatially sorted IDs are not 
required by the code. Second, the software masks for the independent STOMP regions stored in the HDF5 pair 
file assuming the input sub-sample likewise has information on the STOMP regions. For the data we use and 
clustering-$z$s we show in Section \ref{sec:photoz_recoveries}, \textsc{The-wiZZ} will spend of order 10s of 
seconds in calculating clustering-$z$s for scales less than $300$ kpc and of order minutes for larger scales for a 
fixed number of cores. This allows users to compute and re-compute clustering-$z$s for any given sample in a 
tractable amount of time. The software for \textsc{pdf\_maker} can also make use of multiple cores ensuring 
scalability to even larger datasets.

\textsc{The-wiZZ} computes its errors through spatial bootstrapping utilizing the STOMP regions that were 
previously calculated in \textsc{pair\_maker}. Thanks to these independent regions and clever bookkeeping, 
\textsc{The-wiZZ} can compute thousands of bootstrap realizations and calculate errors nearly instantly. 
\textsc{The-wiZZ}  even allows for the storage of intermediate data products such as the over-densities in each 
region and the individual bootstrap samples, allowing one to propagate errors in the clustering redshift estimate 
into any analysis that utilizes the clustering redshift distributions \textsc{The-wiZZ} produces.

An important differentiation between \textsc{pair\_maker} and \textsc{pdf\_maker} is that the latter does not 
require that STOMP be installed or run. \textsc{pdf\_maker} uses very few non-standard \textsc{Python} 
packages and those it does use can be easily installed through \textsc{pip} or come with an installation of the 
popular \textsc{Anaconda}\footnote{\url{http://www.continuum.io/why-anaconda/}} distribution of \textsc{Python}. 
Data products produced from \textsc{The-wiZZ}'s \textsc{pair\_maker} can then be widely distributed as a
value-added catalog product with end users only needing to use \textsc{pdf\_maker} to produce robust
clustering-$z$s. This is the main power of \textsc{The-wiZZ} in enabling science with clustering redshifts. In the 
remainder of the paper we will show how the flexibility of \textsc{The-wiZZ} enables unique uses of clustering 
redshifts, such as producing clustering redshift estimates for individual galaxies using machine learning. 


\subsection{Bias mitigation}\label{sec:bias_mitigation}

Properly mitigating the effect of galaxy bias in clustering redshift estimates is essential to using these redshift 
distributions in any scientific analysis. There is a large amount of literature on this topic and we will not go into 
this in depth as it is not the focus of this article. \textsc{The-wiZZ} does not currently implement a technique for 
mitigating the effect galaxy bias, leaving the choice up to the user. In general galaxy bias mitigation 
techniques can be thought of as a post-processing applied to the output of The-wiZZ, even those of e.g. \citet[][]
{newman08, mcquinn13}. We will point out however that \textsc{The-wiZZ} is perfectly suited for many of the 
literature techniques suggested. One specific example is the technique of \citet{schmidt13} and \citet{menard13} 
which show that by preselecting a narrow redshift range of unknown objects, one can attempt to mitigate the 
effect of galaxy bias. Indeed, \citet{rahman16b} showed that this is the case when selecting Sloan Digital Sky 
Survey (SDSS) galaxies in narrow photo-$z$ bins and summing the individual clustering-$z$s to create the 
clustering-$z$s for magnitude-limited samples. \citet{rahman16a, rahman16b}, and \citet{scottez16} showed that 
one can use selections in galaxy color to achieve a similar effect. \textsc{The-wiZZ} is ideal for producing such 
preselected clustering-$z$s as it enables the redshift estimation of any sub-sample of the unknown galaxy 
sample considered. Since the galaxy bias removal is a post-processing step, function forms of 
the galaxy bias could be provided along with the data files to run \textsc{The-wiZZ} for a given set of data. This 
could be very powerful in enabling science for end users and add to legacy value in the context of a galaxy 
survey.

For the results shown in Section~\ref{sec:photoz_recoveries}, we implement a simplified version of 
these bias mitigations which is similar to that shown in \citet{schmidt13}'s Figure 5 and the "(no bias)" photo-z 
sampling fromm \citet{rahman16b}. This simplified bias mitigation preselects in narrow redshift bins using photo-
zs. For narrow redshift distributions the galaxy bias evolution is close to a constant over the peak of the redshift 
distribution. Clustering-zs using these narrow selections can then be summed together, creating a clusering-z 
measurement for a larger redshift range that has much of the effect of galaxy bias mitigated. This assumes that 
the evolution of the galaxy cross-bias between the unknown and reference samples is smooth and well behaved, 
an assumption that is likely broken when, for instance, the reference sample's selection changes(e.g. switching 
from LRGs to QSOs). This can be thought of as a first order correction to the galaxy bias. Precision cosmology 
measurements will likely need to further mitigate the effects of galaxy bias using the spectroscopic bias evolution 
for example\citep[see][]{rahman16b, scottez16}, however for analyses that require less precision this simplified 
method is an ideal way of using clustering-zs in a straight forward manner. If one does not have access to photo-
zs, color or brightness cuts could also be employed as long as they represent fairly narrow selections in redshift. 
We follow the pre-selection in photo-z bias mitigation technique in Section \ref{sec:photoz_recoveries}.


\section{Data}\label{sec:data}

We use several different sets of reference data and one set of unknown data in demonstrating the capabilities 
of \textsc{The-wiZZ}. The data come from the large spectroscopic catalogs of The Sloan Digital Sky 
Survey(SDSS) and The Galaxy and Mass Assembly survey (GAMA). Throughout this analysis we use 
photometric data with unknown redshifts from The Kilo Degree Survey (KiDS). The data we use is an excellent 
test bed for the \textsc{The-wiZZ}'s ability to scale to future, high data volume surveys.


\subsection{Photometric, unknown data}

The photometric data we use come from KiDS. KiDS represents a large area lensing survey that shows
\textsc{The-wiZZ}'s ability to scale to future datasets such as LSST, Euclid, and WFIRST.


\subsubsection{The Kilo-Degree Survey (KiDS)}

The ongoing Kilo Degree Survey\footnote{\url{http://kids.strw.leidenuniv.nl/}} \citep[KiDS,][]{dejong15} is a 
$1\,500~\rm{deg}^2$ survey observed with OmegaCAM on the VLT Survey Telescope (VST) in SDSS-like $u-$,  
$g-$, $r-$, $i-$bands down to $5 \sigma$ limiting magnitudes of 24.3, 25.1, 24.9, and 23.8 AB, respectively. 
The survey is designed for weak lensing and has a median seeing of better than 0.7" in the $r-$ band. Further 
details on the survey can be found in \citet{dejong15}, \citet{kuijken15} and \citet{hildebrandt16}. For this 
analysis we use catalogs and automated masks of bright stars and image defects produced by
\textsc{Astro-WISE} \citep{valentijn07, begeman13} and THELI \citep{erben05, schirmer13}. Magnitudes and 
colors are produced using \textsc{GAaP}, a seeing Gaussianization process that produces consistent aperture 
photometry across the different observed bands \citep{kuijken08, kuijken15}. Initial detection catalogs for 
photometry use \textsc{SExtractor} \citep{sextractor}. For photometric redshifts we use a modified version of 
the Bayesian Photometric Redshifts \citep[\textsc{BPZ}]{bpz} code as described in \citet{hildebrandt12} and 
\citet{hildebrandt16}.

The data we use from KiDS are a currently non-public, early data product dubbed KiDS-450. This iteration of the 
survey has an area of roughly $450~\rm{deg}^2$ and covers all GAMA fields in all four KiDS bands. The survey 
is also covered by spectra from the north Galactic cap of The Sloan Digital Sky Survey. After applying the full 
masking that intersects with the northernly GAMA and SDSS fields we have a total area of $\sim170~\rm{deg}
^2$. In this analysis we mimic the cuts described in \citet{hildebrandt16} for comparison to the redshift 
distributions shown therein. We utilize the shape weights produced by \emph{Lens}fit \citep{miller13, 
fenechconti16} as weights for each object to further mimic this selection. These weights also act as a magnitude 
limit, returning low and zero weights for galaxies with $r>25$. The cuts we make also exclude all galaxies with 
$r<20$. In total our sample of photometric, unknown objects is 3\,959\,558 total galaxies with weight $>0$.We 
make a further cut for the analysis we present in Section \ref{sec:kdtree}, additionally requiring 
that the \textsc{GAaP} magnitude in each band has a value of $>0$ assuring that each magnitude is observed 
(not necessarily detected) for each object. For this sample, \textsc{GAaP} values of $99$, defined as
non-detected in any band are replaced with the limiting magnitude in that band.


\subsection{Spectroscopic, reference data}

\begin{figure}
    \includegraphics[width=0.495\textwidth]{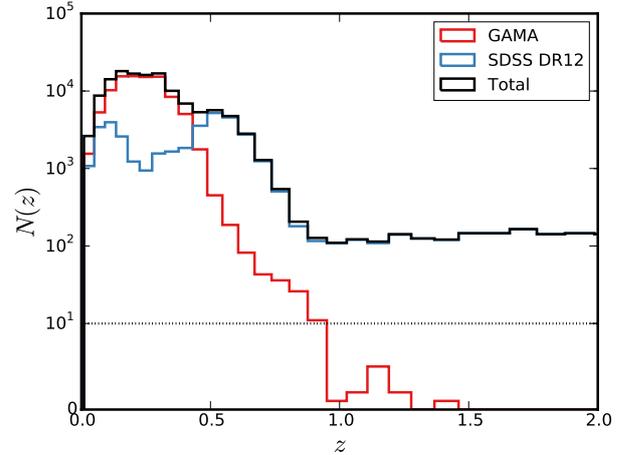}
    \caption{
    \label{fig:spec_dist}
    Symmetric-log plot of the number of spectra overlapping with the current KiDS coverage as a function of 
    redshift. The GAMA survey is the dominant sample for low redshifts with SDSS dominating for high redshift. 
    The large amount of objects above $z=1.0$ are spectroscopic QSOs. The data are binned linearly in $
    \ln(1+z)$ and follow the exact binning that we will later use in our clustering redshifts. Above the dotted line 
    the data are plotted logarithmically in $N(z)$, below they are plotted linearly. For this plot we show galaxies in 
    the GAMA catalog that have spectral redshifts from SDSS as SDSS galaxies.
    }
\end{figure}

We use two spectroscopic surveys for our reference sample, the GAMA survey and SDSS Data Release 
12 (DR12). The distribution as a function of redshift for both surveys within the overlapping area of KiDS is 
shown in Figure \ref{fig:spec_dist}. As seen in the figure, GAMA dominates for low redshift while SDSS 
dominates a higher redshift. In total there are 135\,567 galaxies in the sample we use spanning a redshift range 
of $0.01<z<7.0$. At redshifts $z>1.0$ we rely almost exclusively on spectroscopic Quasi-Stellar 
Objects (QSOs) from the SDSS DR12 catalog. In addition to using the quality cuts provided by each survey, we 
also reject all spectra within a 2 arc-second radius of each other to remove duplicate objects.

\subsubsection{Galaxy and Mass Assembly (GAMA) survey}

We make use of non-public spectroscopic data, dubbed GAMA-II, from the GAMA survey\footnote{\url{http://
www.gama-survey.org/}} \citep{driver09, baldry14, liske15}. GAMA is a magnitude limited spectroscopic survey 
covering over $286~\rm{deg}^2$ on the Anglo-Australian Telescope (AAT) using the AAOmega multi-object 
spectrograph. The GAMA survey is designed to study galaxy and mass evolution at low and intermediate 
redshifts, however the spectra can also be used in cross-correlation studies such as this one. For our analysis 
we make use of the equatorial fields of GAMA that overlap with KiDS dubbed G09, G12, and G15 corresponding 
to their RA center. These fields are primarily observed to a limiting magnitude of $r<19.8$ over
$180~\rm{deg}^2$. We select spectroscopic redshifts from the survey that satisfy their "Main Sample" criteria 
(SURVEY\_CLASS $\geq3$) and have a redshift quality value of $\rm{nQ} \geq 3$.GAMA contains galaxies from 
other spectroscopic surveys including SDSS to reach its level of completeness. We reject any galaxy from the 
SDSS catalog that is within 2 arc-seconds of a GAMA catalog galaxy to avoid duplicate redshifts between the 
two catalogs. We make use of the GAMA redshift completeness masks to exclude bad area from the survey and 
limit the area we must search for pairs in. After masking for the KiDS and GAMA combined area we have $101~
\rm{deg}^2$ with a total number of 94\,694 unique spectroscopic galaxies from the GAMA catalog.


\subsubsection{Sloan Digital Sky Survey (SDSS) DR12}

We make use of spectra from the 12th data release from the Sloan Digital Sky Survey\footnote{\url{http://
www.sdss.org/}} \citep{york00, eisenstein11, alam15}. SDSS not only adds low- and intermediate-redshift 
spectra but also spectroscopic QSOs that allow us to produce clustering-$z$s out to very high redshift. For our 
purposes we make use of all galaxy spectra from the survey that overlap with KiDS. This nominally includes 
galaxies from the SDSS main sample \citep{strauss02}, Baryon Oscillation Spectroscopic Survey \citep[BOSS,]
[]{dawson13} galaxies both from the LOWZ and CMASS samples, and the aforementioned QSOs \citep{ross12}. 
The galaxies we utilize are those defined as ``Science Quality'' from the SkyServer catalog and have a redshift 
quality selection with $\rm{zWarning} = 0$ \citep{bolton12}. As stated previously, objects are also checked for 
duplication between SDSS and GAMA. The mask we use for this analysis comes from converting the SDSS 
Mangle polygons into STOMP format. This STOMP Map was previously used in the analyses of 
\citet{schmidt15}, \citet{rahman15}, and \citet{rahman16b}. The overlapping area between the current coverage 
of KiDS andSDSS/BOSS is $170~\rm{deg}^2$ containing 40\,873 objects in total.


\section{Clustering redshifts}\label{sec:recoveries}

\begin{figure*}
    \includegraphics[width=1.0\textwidth]{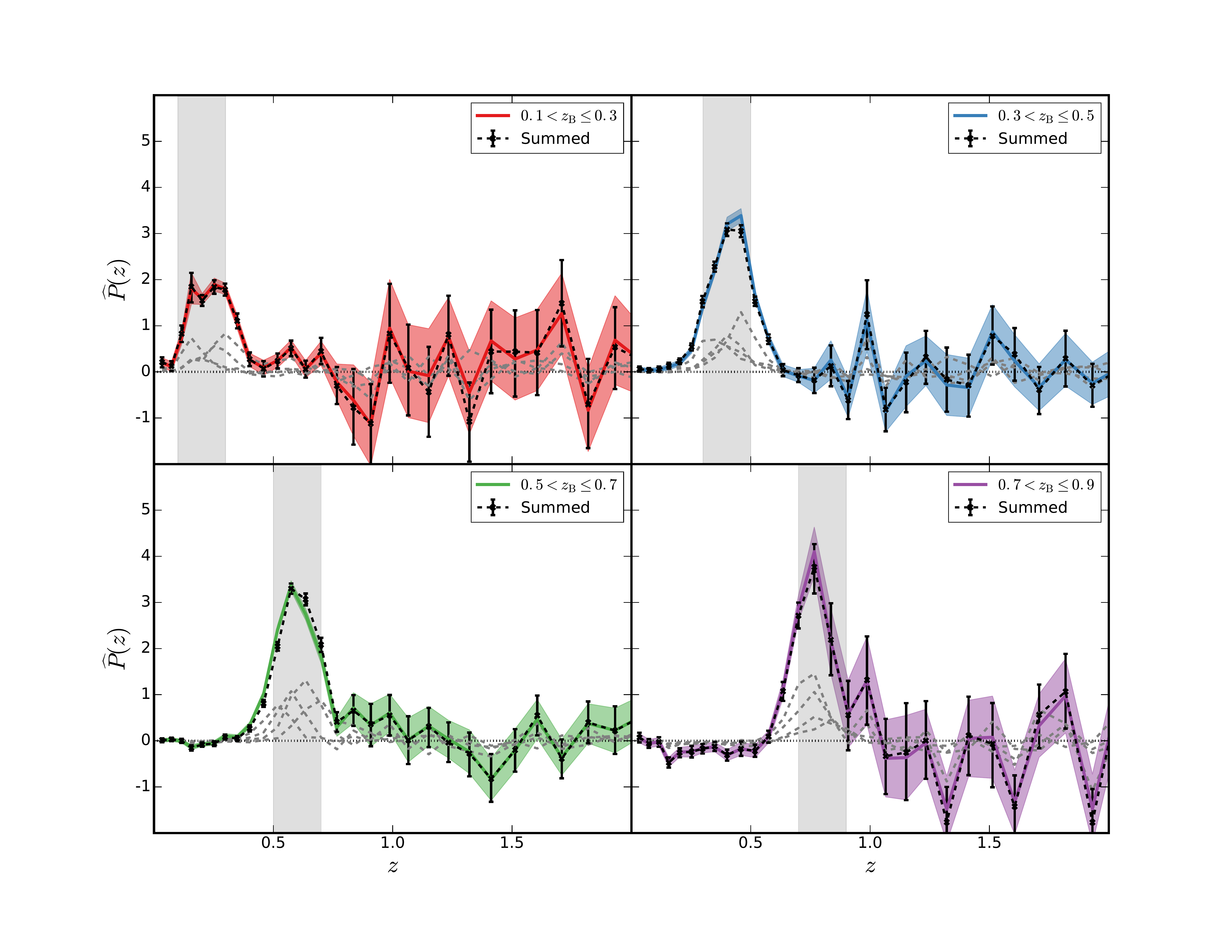}
    \caption{\label{fig:recovery_bins}
    Raw and summed clustering-$z$s produced by \textsc{The-wiZZ} using objects from KiDS selected in 
    $z_{\rm B}$ as the unknown sample and GAMA and SDSS spectra as the reference sample normalized into 
    an estimated PDF. Colored bands are clustering-$z$s from selections in $z_{\rm B}$ mimicking the bins of 
    \citet{hildebrandt16} (CS bins). The light grey regions show the selection in photo-$z$. Grey dashed lines are 
    the cluster-$z$s produced by dividing the CS bins into 4 sub-bins with $\Delta z_{\rm B}=0.05$ normalized by 
    their number of objects relative to the CS bin. The grey dashed lines appear to sum up to the CS bin as 
    a function of redshift suggesting the galaxy bias in the clustering redshift estimate is well behaved. Black data 
    points are the resultant clustering-$z$ from normalizing, summing, and averaging the individual spatial 
    bootstraps of the sub-bins into the full CS bin. The bins were all selected from the same catalog and use the 
    same \textsc{The-wiZZ} data file demonstrating how clustering-$z$s can be quickly created for a variety of 
    samples using \textsc{The-wiZZ}.
    }
\end{figure*}

In this section we show clustering-$z$s produced by \textsc{The-wiZZ} from various sub-samples of KiDS as the 
unknown samples with the reference data coming from SDSS and GAMA. This is not the first time a clustering 
redshift technique has been applied to the KiDS data, with lower redshift and smaller galaxy sample results 
shown in \citet{hildebrandt16} and \citet{johnson16} using different clustering redshift estimators and wider 
binning in photo-$z$. We compare to the results of \citet{hildebrandt16} in Section \ref{sec:discussion_kids_cs}.

We show the power of \textsc{The-wiZZ} in producing clustering-$z$s using the same catalogs and data 
files. The clustering-$z$s for a given unknown data sample are all produced from the same data, we only 
change how we select the given sub-samples used. Throughout this section the \textsc{pair\_finder} portion of 
the software is only run once. We start in Section \ref{sec:photoz_recoveries} by producing clustering 
redshifts using photometric redshift peak probability ($z_{\rm B}$) as a pre-selection as suggested in 
\citet{menard13} and \citet{schmidt13} and shown in \citet{rahman16b} and \citet{scottez16}. Then in Section 
\ref{sec:kdtree}, we introduce a novel technique where we estimate the redshifts of individual objects using a
$k$-dimensional spatial search tree (\textsc{kdTree}) based method that allows us to select the $k$-nearest 
neighbors to an object in color-magnitude space, run the software on those neighbors, and produce a clustering 
redshift estimate for individual galaxies. For all the analyses we present in this section we use the 
same randoms for use in the natural estimator. These are drawn to have a size of 10 times the total 
photometric sample. This means that every sub-sample has a large number of randoms compared to the number 
of objects used in the sub-sample.

We estimate errors and covariances by spatially bootstrapping 1000 times over 279 independent spatial regions 
as defined by STOMP. The spatial regions approximate the individual $1~\rm{deg}^2$ pointings from KiDS. We 
also separate the analysis into two parts: computing the over-densities in regions where SDSS and GAMA both 
overlap KiDS (representing 166 regions) and the regions where only SDSS overlaps KIDS (113 regions). We 
then combine these regions by spatial bootstrap which smoothly joins the two surveys' redshift overlap. 
Throughout this analysis we measure the cross-correlation amplitude on physical scales between $R=0.1-1.0\, 
\rm{Mpc}$. We bin the reference galaxies in redshift with 50 bins equally spaced in $\ln(1+z)$ from 
$0.01<z<6.0$. We only plot and normalize the data to a redshift of $z=2.0$ for clarity and to compare to the 
redshift distributions from \citet{hildebrandt16}. Because of measurement noise, spatially 
dependent survey systematics, and changes in unknown galaxy selection function, some points in the 
clustering-$z$s are negative. We treat these negative points by inverse variance averaging them with 
neighboring bins until all points are positive definite. Without this smoothing, negative points with 
large values and error bars will bias the norm. This is also true of points with large positive values and error 
bars. This allows us to properly convert the clustering-$z$s into PDF estimates assuming the bias is well 
mitigated. We compute the normalization using this adaptive smoothing but plot the data as measured. 
Normalizations to transform the over-densities into an estimated redshift PDF are computed using a trapezoidal 
sum with fixed end points of $z=0.01,2.0$. 


\subsection{Photo-$z$ selection}\label{sec:photoz_recoveries}

\citet{menard13} and \citet{schmidt13} demonstrate that one way to mitigate the effect of galaxy bias in 
clustering redshifts is to utilize color or photometric redshift information to preselect a sample of galaxies in a 
narrow redshift range, making the galaxy bias as constant as possible. One can then add clustering-$z$s of 
these preselected samples together using their relative numbers to produce the redshift distribution for a larger 
sample where the effect of the bias has been mitigated. \citet{rahman16b} and \citet{scottez16} apply this 
method to real data from SDSS and the Canada-France-Hawaii Telescope Legacy Survey (CFHTLS) 
respectively and show that indeed the bias is mitigated by using these narrow redshift samples. This technique 
works best when the resultant redshift distributions are singularly peaked and narrow. If the distribution is found 
to have long tails in $z$ or is multiple peaked, the galaxy bias mitigation will not be as robust.

The design of \textsc{The-wiZZ} enables this kind of clustering-$z$ and bias mitigation very simply.
Sub-samples can be selected and re-selected without having to re-run any correlations, significantly increasing 
the ease at which this method can be applied. We apply a simplified version of the bias mitigation of 
the previously mentioned clustering-z analyses which we describe in  in detail in 
Section~\ref{sec:bias_mitigation}. We attempt to recreate the photometric redshift distributions from the 
KiDS-450 cosmic shear (here after referred to as the CS bins) results \citep{hildebrandt16} as a test of 
\textsc{The-wiZZ}. This is a sample of 4 redshift bins with a width of $\Delta z_{\rm B} = 0.2$, spanning the 
range of $0.1< z_{\rm B} \leq 0.9$ selected by the peak of the redshift posterior, $z_{\rm B}$. We further divide 
each of these bins into 4 smaller photo-$z$ sub-bins with a width of $\Delta z_{\rm B} = 0.05$. This selection is 
pushing the limits of the errors of the photo-$z$s which are similar in size or slightly larger than $\Delta z_{\rm B} 
= 0.05$ for some redshifts.

Figure \ref{fig:recovery_bins} shows the clustering redshifts produced by running \textsc{The-wiZZ} on the CS 
bins plotted as colored bands. The CS bins are normalized to a sum of $1$ over the range $z=0.01-2.0$. The 
clustering-$z$ of the CS bin agrees largely with previous results from \citet{hildebrandt16}, especially the 
detection of a significant tail to high-redshift in the $0.1<z_{\rm B}\leq0.3$ bin. In addition the high redshift bins 
appear to be largely free of low-redshift interlopers, which is again in agreement with the KiDS-450 results. We 
show a direct comparison to the distributions in \citet{hildebrandt16} in Section \ref{sec:discussion_kids_cs}. 
The clustering-$z$s also agree in overall shape and peak position compared to the previous results. The
sub-bins, shown as grey dashed lines, are normalized to the number of objects they contain relative to their 
corresponding CS bin in addition to sum normalization. These bins are single peaked, normalizable, and appear 
to properly sum to the full CS bin. This suggests that the galaxy bias is already fairly constant across the redshift 
range shown. If the bias were evolving strongly with redshift for either the reference or unknown sample one 
would likely see discrepancies between the sub-bins and CS bins.

\begin{figure}
    \includegraphics[width=0.5\textwidth]{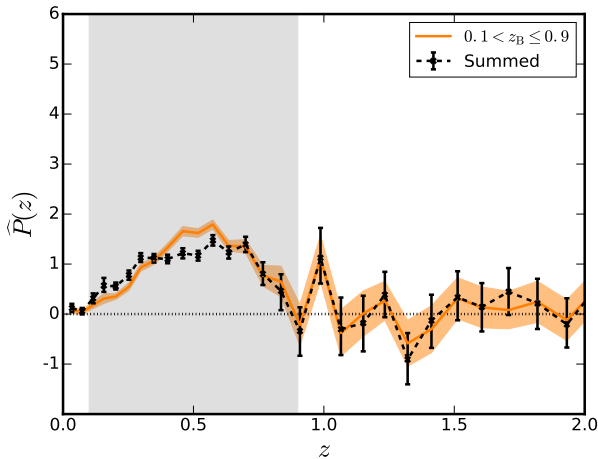}
    \caption{\label{fig:total_bin}
    Raw and summed clustering-$z$s for the total, spanning bin of $0.1<z_{\rm B}\leq0.9$. The orange 
    colored band shows the clustering-$z$ result from running \textsc{The-wiZZ} on the full $0.1<z_{\rm B}
    \leq0.9$ sample. The black data points are the clustering-$z$ created from summing the spatial bootstraps of 
    the clustering-$z$s from the 16, $\Delta z_{\rm B}=0.05$ sub-bins into the total bin. The light-grey region 
    shows the $z_{\rm B}$ selection. We sum the clustering-$z$s together in this manner to mitigate galaxy bias 
    in the clustering-$z$s as suggested in \citet{menard13} and \citet{schmidt13}. These works show how 
    clustering-$z$s of wide distributions in $z$ are much more susceptible to the effect of galaxy bias than 
    narrow selections. The difference between the raw and summed clustering-$z$s shows the effect of this bias 
    mitigation. The design of \textsc{The-wiZZ} is well suited for bias mitigation strategies such as this.
    }
\end{figure}

In order to add the sub-bins together to create the CS bins as well as the larger $0.1< z_{\rm B} \leq 0.9$, total 
spanning bin we make use of the 1000 bootstrap samples. First we ensure that the spatial bootstraps we draw 
for each sub-sample are the same so we can compute proper errors and covariances both within each CS bin 
we create from the summed sub-bins and also covariances between the summed CS bins. We first compute one 
normalization from the average of the spatial bootstraps for each sub-bin. This is due to each bootstrap 
realization being too noisy to properly compute a normalization. We compute all of the normalizations in a range 
$z=0.01-2.0$ except for the $0.7<z_{\rm B}<0.9$. Here we use $z=0.3-2.0$ as our normalization range due to 
the significant low redshift negative peak causing the computation to not converge. The redshift distributions of 
\citet{hildebrandt16} show no significant amplitude at $z<0.3$ so we don't expect this cut to bias our results 
significantly. We apply these sub-bin normalizations to each of the sub-bins' spatial bootstraps and also multiply 
by the number of galaxies in each sub-bin sample for each spatial bootstrap. We then sum these bootstraps 
together to create a new set of 1000 spatial bootstraps for each summed CS and total bin. We then compute the 
median, low side and high side errors by calculating the 16th, 50th, and 84th percentiles from the spatial 
bootstraps. We do this as the percentiles are much more stable than the simple mean and variance. We also 
calculate the mean and median of each of the redshift bins. The mean is calculated the same as the 
normalization using a trapezoidal sum while the median is taken as the point where the cumulative density 
function (CDF) has a value of 50\%. We compute mean and median on the averaged, positive definite 
clustering-$z$s for each bootstrap and then compute the same percentiles as mentioned previously for central 
values and errors.

\begin{center}
\begin{table*}
{
\caption{Properties and summary statistics of the clustering redshift estimates for the KiDS-450 cosmic shear 
              sample using both the clustering-$z$ from running on the cosmic shear sample directly ($z_{\rm CS}$) 
              and after summing the clustering-$z$ sub-bins to from the CS sample ($z_{\rm sum}$). 
              $z_{\rm HH}$ are the median and mean results from the spectroscopic re-weighting
              scheme presented by \citet{hildebrandt16} in their Table 1 as $z_{\rm DIR}$.}
\label{tab:recovery_summary}
\small
\hfill{}
\begin{tabular}{c c c c c c c c c}
    \hline
    bin & $z_{\rm B}$ range & no.of unknown objects & median($z_{\rm{CS}}$) & $\langle z_{\rm{CS}}\rangle$ &
    median($z_{\rm{sum}}$) & $\langle z_{\rm{sum}}\rangle$ & median($z_{\rm HH}$) & $\langle z_{\rm HH} 
    \rangle$\\ \hline \smallskip
    1 & $0.1<z_{\rm B}\leq0.3$ & 1\,199\,854 & $0.596^{+0.462}_{-0.258}$ & $0.751^{+0.176}_{-0.366}$ & 
    $0.357^{+0.574}_{-0.062}$ & $0.698^{+0.308}_{-0.354}$ & $0.418\pm0.041$ & $0.736\pm0.036$ \\ \smallskip
    2 & $0.3<z_{\rm B}\leq0.5$ & 940\,381 & $0.464^{+0.032}_{-0.015}$ & $0.488^{+0.058}_{-0.025}$ &
    $0.458^{+0.022}_{-0.022}$ & $0.473^{+0.218}_{-0.023}$ & $0.451\pm0.012$ & $0.574\pm0.016$ \\ \smallskip
    3 & $0.5<z_{\rm B}\leq0.7$ & 951\,747 & $0.633^{+0.018}_{-0.015}$ & $0.683^{+0.138}_{-0.048}$ &
    $0.649^{+0.028}_{-0.018}$ & $0.670^{+0.117}_{-0.049}$ & $0.659\pm0.003$ & $0.728\pm0.010$ \\ \smallskip
    4 & $0.7<z_{\rm B}\leq0.9$ & 867\,576 & $1.255^{+0.372}_{-0.363}$ & $0.969^{+0.313}_{-0.264}$ &
    $1.276^{+0.310}_{-0.300}$ & $0.985^{+0.269}_{-0.214}$ & $0.829\pm0.004$ & $0.867\pm0.006$ \\ \smallskip
    1-4 & $0.1<z_{\rm B}\leq0.9$ & 3\,959\,558 & $0.604^{+0.037}_{-0.037}$ & $0.704^{+0.113}_{-0.074}$ &
    $0.606^{+0.051}_{-0.044}$ & $0.643^{+0.155}_{-0.071}$ & - & - \\ \smallskip
\end{tabular}}
\hfill{}
\end{table*}
\end{center}

The black data points in Figure \ref{fig:recovery_bins} show the results of this process for each of the 4 CS 
summed bins. The summed data have slightly larger error bars than that of the CS clustering-$z$s largely due 
to the extra normalization step during the addition. The shapes of the clustering-$z$s between the summed and 
CS clustering-$z$s are similar but there are slight differences. These differences mainly show up in the 
$0.3<z_{\rm B}\leq0.5$ bin. This bin finds its peak slightly shifted to higher redshift relative to the CS 
clustering-$z$. There are also slight differences in the peak of each of the other bins. We show the total, 
$z_{\rm B}$ spanning bin in Figure \ref{fig:total_bin}. This bin has its low redshift amplitude increased and 
intermediate redshifts suppressed in the summed clustering-$z$ relative to the raw clustering-$z$. If one 
assumes that the galaxy bias is increasing with redshift for both the reference and unknown samples this is the 
trend one should expect as the bias exaggerates the amplitude of the clustering-$z$s at high redshift compared 
to the truth. This is also reinforced by the fact that the peak position of the redshift bins with narrower 
distributions are largely unchanged between the summed and CS clustering-$z$s. The increase in 
noise at higher redshift is likely two fold: First there are many fewer reference galaxies at these redshifts; Second 
the increase in reference bias accentuates any marginal correlation at these redshifts, increasing the noise. The 
later is partially mitigated by the narrow redshift bins but would likely require explicit removal for the reference 
bias to be completely accounted for.

Table \ref{tab:recovery_summary} summarizes the single point statistics we measure for the original and 
summed CS clustering-$z$s. For the majority of bins, both the mean and median of the clustering-$z$s are 
consistent within their error bars between the raw CS clustering-$z$ and the summed version. 


\subsection{Color selection with \textsc{kdTrees}}\label{sec:kdtree}

We can also utilize the colors of the unknown sample objects themselves to determine a mapping from color to 
redshift rather than relying on photo-$z$ to make the mapping for us. Such selections have been carried out in 
\citet{menard13}, \citet{rahman16a}, \citet{rahman16b}, and \citet{scottez16} to select narrow distributions in 
color and thus narrow distributions in redshift. We can then compute the clustering redshifts of these color 
selected samples for a color-color cell and create an estimate for most sub-samples of galaxies by assigning 
them to a cell.

\begin{figure}
    \includegraphics[width=0.5\textwidth]{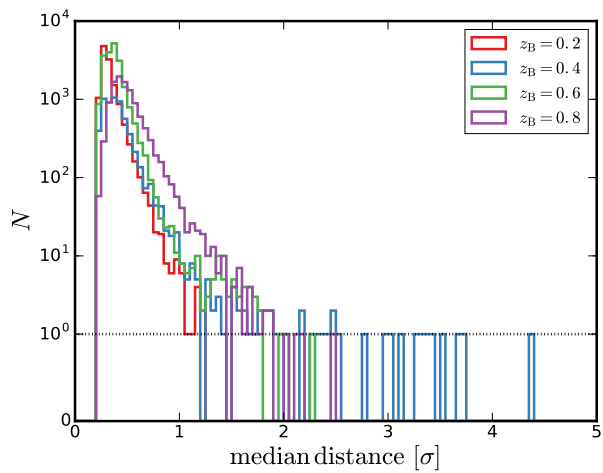}
    \caption{\label{fig:kdtree_median_distance}
    Symmetric-log plot of the histograms of the median distance in normalized units of the 4\,096 
    objects matched to the input object from the \textsc{kdTree}. The different curves are for each of the
    photo-$z$ selected galaxy samples we use to test the method. That most of the median distances are below 
    a value of $1\sigma$ away in color-magnitude space lends credence that the self-similar galaxies selected 
    from the \textsc{kdTree} are representative of the input galaxy. Above the dotted line the data are plotted 
    logarithmically, below they are plotted linearly.
    }
\end{figure}

\begin{figure*}
    \includegraphics[width=1.0\textwidth]{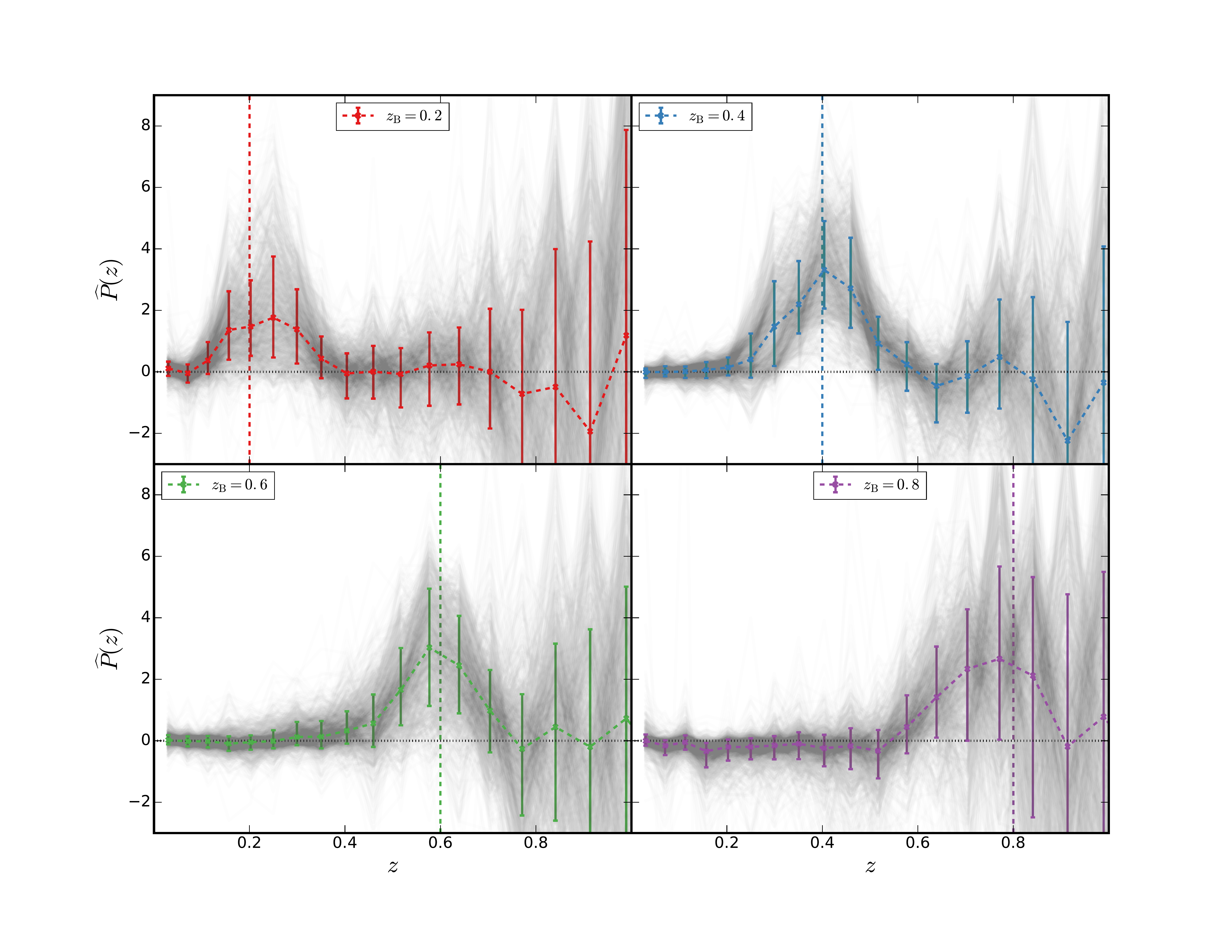}
    \caption{\label{fig:kdtree_bins}
    Single galaxy clustering redshift estimates for each of the samples we compared against. The grey lines are 
    the the individual clustering-$z$s produced by finding self-similar galaxies in color magnitude space for a 
    random sub-sample of input objects from the photo-$z$ selection. The colored data points are the
    median$\pm34\%$ dispersion of the full sample of input object clustering-$z$s. We use photo-$z$
    $z_{\rm B}$ to select a test sample but create the \textsc{kdTree} and self-similar galaxies using the full 
    catalog using no explicit redshift information. The peak redshift found by this method agrees with the
    photo-$z$ estimate validating this hybrid machine learning-cluster-$z$ approach.
    }
\end{figure*}

A major failing of this method is the high dimensional partitioning of color-color space and efficiently 
populating each partition. A way around this is to use machine learning to reduce the dimensionality 
of the problem. This dimensionality reduction allows for the use of any galaxy quantity that correlates with 
redshift to arbitrary complexity within the limits of the machine learning algorithm chosen. We chose a relatively 
simple method by creating a \textsc{kdTree} in a space defined by several galaxy properties. This is similar to 
color-space re-weighting techniques for photo-$z$s described in \citet{lima08} and \citet{cunha09}. By using this 
\textsc{kdTree} we can create clustering redshifts for individual galaxies by matching a single galaxy into the 
\textsc{kdTree} and measuring a clustering-$z$ on the self-similar objects that the \textsc{kdTree} returns. This 
method can be extremely useful for survey users interested in individual or small samples of unique galaxies. It 
is also possible to use this cluster-$z$ as a prior for Bayesian based photo-$z$ methods. This will also be  useful 
for predicting clustering-$z$s in surveys that are observed with similar band pass filters but contain little or no 
spectroscopic overlap by matching their objects into the survey with spectra.

For this analysis we use a sub-set of the full catalog, limiting ourselves to the KiDS area intersecting both SDSS 
and GAMA. We also ensure that the objects are ``observed'' in each band that is the \textsc{GAaP} magnitude 
returns $>0$. In total roughly 2.8 million galaxies remain in this sample compared to the 4 million previously 
used. We use the 3 \textsc{GAaP} colors $u-g,g-r,r-i$ as well as the $r$ band magnitude as the space to create 
our \textsc{kdTree} in. We treat these colors and magnitudes similarly to that of BPZ where non-detections in a 
given band are replaced with the limiting aperture magnitude in the appropriate band. We create the 4 
dimensional \textsc{kdTree} after we standardize the colors and $r$ magnitude to have mean of 0 and variance 
of 1. This regularizes the tree and prevents dimensions with large variance from dominating the Euclidean 
distances and therefore the computed neighbors. We make use of the package \textsc{cKDTree} from 
\textsc{scipy}\footnote{\url{http://www.scipy.org/}} to create the \textsc{kdTree}.

For each unknown object we then match the same properties we created the \textsc{kdTree} with into the tree 
and then return the nearest 4\,096 unknown objects with similar properties as identified by the \textsc{kdTree}. 
Figure \ref{fig:kdtree_median_distance} shows the median distance of the 4\,096 objects to each input, matched 
object in color-magnitude space for the samples we consider. This data can be used as a quality statistic, 
removing objects that were matched with large distances relative to the rest of objects matched in. The 
distances plotted in this figure are plotted in standard deviations relative to the normalized color-magnitude 
distributions. It is likely that objects matched with a median distance of larger than $1\sigma$ are not well 
represented in color-magnitude space and will produce inaccurate clustering-$z$s. We can then input these
4\,096 objects into \textsc{The-wiZZ}to produce a clustering redshift estimate for the individual object we 
matched into the \textsc{kdTree}. We select 4\,096 objects as it gives us relatively stable statistics per STOMP 
region ($N\sim24$) and is not so wide as to have too many of the median distances beyond $1\sigma$ in
color-magnitude space.

To test our method we first select galaxies with photo-$z$ values of $z_{\rm B}=0.2,0.4,0.6,0.8$, the midpoints 
of the CS bins. We make this selection to have a rough idea of what the redshift is before creating the 
clustering-$z$s for comparison. The \textsc{kdTree} is created from the full catalog and has no direct knowledge 
of redshift. We create single galaxy clustering-$z$s for each of the $12\,654$, $5\,968$, $20\,292$, $10\,882$ 
galaxies in each sample respectively. Figure \ref{fig:kdtree_bins} shows the individual clustering-$z$s for each 
galaxy in these samples as grey lines. Darker regions represent redshifts where the clustering-$z$s are similar. 
The data points and bars shown are the median$\pm34\%$ showing the dispersion of the individual clustering-
$z$s to give a sense of how the clustering-$z$s are distributed. For clarity, we normalize the individual 
distributions to a redshift range of $z=0.01-1.0$ when plotting. The relatively few unknown objects we use and 
the few high redshift QSOs in this footprint prevents interpretation of the clustering-$z$s beyond $z=1.0$.


\section{Discussion}\label{sec:discussion}

\begin{figure*}
    \includegraphics[width=1.0\textwidth]{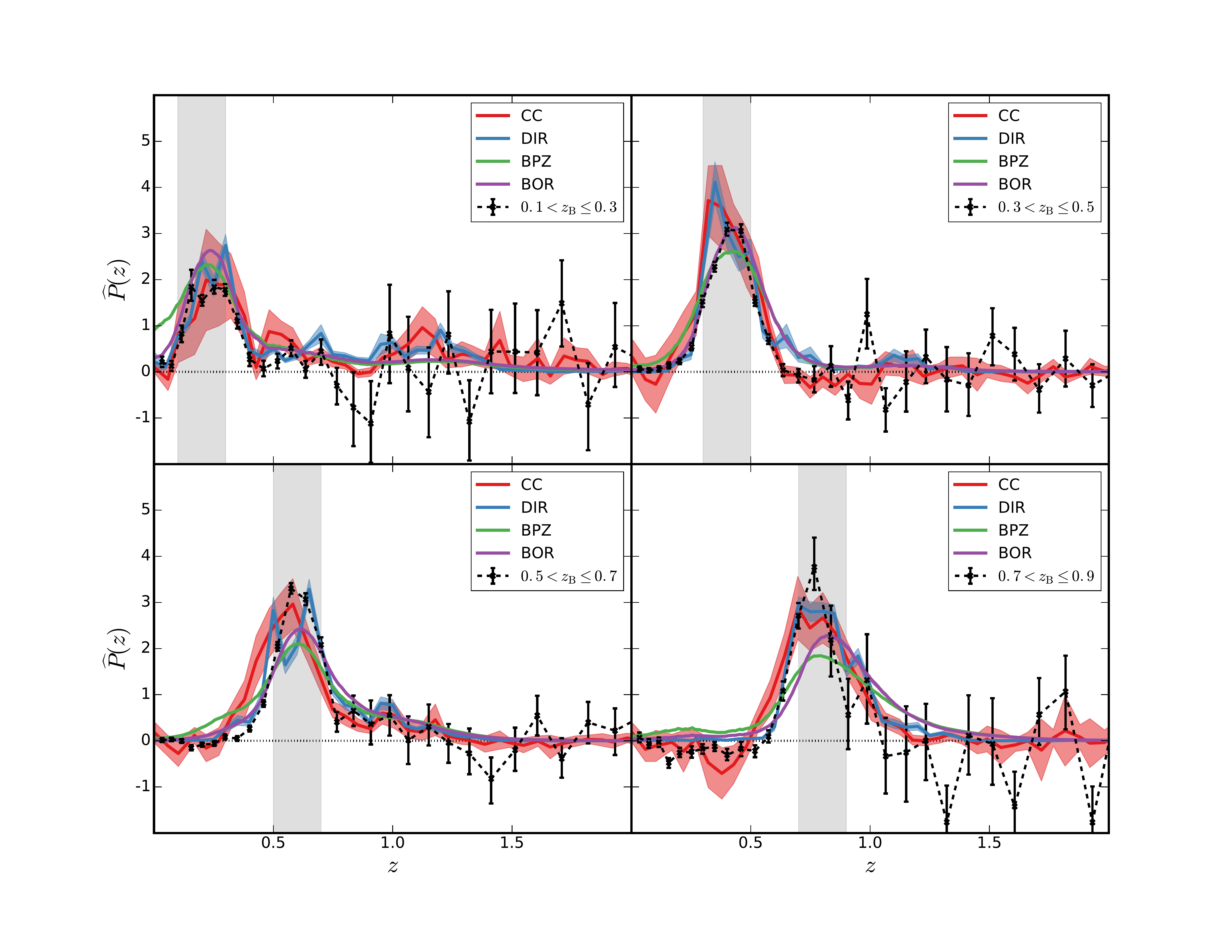}
    \caption{
    \label{fig:kids_cs_comparison}
    Comparison of the summed CS bins with the redshift distributions of \citet{hildebrandt16}. CC (red) refers to 
    the cross-correlation technique used in that work with spectroscopic galaxies coming from the zCOSMOS 
    \citep{lilly09} and DEEP2 \citep{newman13} redshift surveys, BPZ (green) refers to the summed posteriors 
    from BPZ, DIR (blue) refers to the direct recalibration scheme described in that work, and BOR (purple) refers 
    to the posteriors re-weighted as described in \citet{bordoloi10}. Overall the peak position and amplitudes 
    agree with the fiducial method of the KiDS cosmic shear sample (DIR).
    }
\end{figure*}

The clustering redshifts shown in the previous section demonstrate the flexibility of \textsc{The-wiZZ} in 
producing clustering-$z$s for any sub-sample of the data without having to re-run any 2-point correlations. In 
this section we will discuss our results specifically in the context of the redshift distributions used in 
\citet{hildebrandt16} and some non-cosmology applications.


\subsection{Comparison to KiDS-450 cosmic shear}\label{sec:discussion_kids_cs}

In Figure \ref{fig:kids_cs_comparison} we directly compare the data points from the KiDS-450 cosmic shear 
results with those of clustering-$z$s from \textsc{The-wiZZ}. Overall, the clustering-$z$ distributions presented 
here confirm the distributions shown in \citet{hildebrandt16}. The canonical redshift distribution from KiDS-450, 
the $z_{\rm DIR}$ distribution which reweighs spectroscopic galaxies in color space to account for 
non-representative spectra in calibrating photo-$z$s, agrees with our results with some allowance for sample 
variance. In addition, a previous, pre-\textsc{The-wiZZ} clustering-$z$ code (CC in Figure
\ref{fig:kids_cs_comparison}) also confirms our redshift distribution. This CC clustering-$z$ was created using 
using only $1.6~\rm{deg}^2$ of spectra from zCOSMOS \citep{lilly09} and DEEP2 \citep{newman13} and uses 
the Newman iteration \citep{newman08} to mitigate bias suggesting again that the bias of the clustering-$z$s 
presented here is well behaved.

When comparing the mean and median redshifts of the summed clustering-$z$s to the redshift estimates from 
the direct calibration, $z_{\rm DIR}$, we find gross agreement between the two methods. Comparing to 
\citet[Table 1]{hildebrandt16}, we find that both the means and medians are consistent to within 1 or 2$\sigma$ 
when summing the square of the admittedly large errors. These errors come largely from uncertainty in the 
amplitude of the high redshift tail in the clustering-$z$s. There are only 100 reference spectra per redshift bin for 
this part of the clustering-$z$ leading to the large errors. There are several problems with the clustering-$z$s 
presented which we discuss here.

Around $z=1.0$ there is a feature in the clustering redshifts that shows up in each of the clustering-$z$s shown 
in Figure \ref{fig:recovery_bins}. Given its position in redshift, this feature likely comes from the switch between 
SDSS galaxies and QSOs. GAMA is also not contributing any more galaxies at this point as seen in Figure 
\ref{fig:spec_dist}. This negative correlation could be suppressing the measurement of the high-redshift tail that 
is seen in the $z_{\rm DIR}$ method of \citet{hildebrandt16}. These negative correlations were also seen in the 
cross-correlation technique used in that paper and seem to be a feature of the data rather than a failing of 
\textsc{The-wiZZ}. Negative amplitudes can be caused by incorrect masking or by extremely dense large-scale 
structure as shown in \citet{rahman15}. We did not account for these over-densities by "cleaning" in 
this work. \citet{rahman15} show that this cleaning is required to remove excess positive correlation and excess 
noise at redshifts $z<0.2$. With \textsc{The-wiZZ} however we do not observe such excess correlation when 
comparing it to the software of \citet{rahman15} when using the data. This could be due to the the code of 
\citet{rahman15} using signal matched filter weights in $\theta$ rather than $R_{\rm physical}$. This adds an 
extra factor in of the angular diameter distance in the signal matched weights \citet{rahman15} and similar codes 
use that may accentuate over-densities at low redshift causing these excess amplitudes. When necessary, 
however, the \citet{rahman15} cleaning step can simple be thought of as a pre-processing step on the reference 
sample before it is input into \textsc{The-wiZZ}. It does not require any change to the algorithm.

Another way to cause these negative correlations is if the normalization of the area and randoms are slightly off. 
The density of galaxies in KiDS changes from pointing to pointing largely due to variations in average seeing. We 
try to account for these density variations by using the STOMP regions however these regions are not perfectly 
matched to the pointings and as such could cause the computation of the average density over the survey to be 
incorrect, leading to negative correlations. This could be fixed in the future by accounting for such systematics in 
a way similar to that of \citet{morrison15} or \citet{leistedt15}. These methods produce weight maps 
that can be used to weight the unknown objects similar to how we use shape weights in this analysis. Such 
weight maps should be used for high precision analyses to account for selection effects.

Small discrepancies between these results and \citet{hildebrandt16} could also come from the galaxy bias not 
being completely removed from the samples. The lowest CS redshift bin shows a significant second peak in 
the redshift distribution around $z=0.5$. \citet{schmidt13} shows specifically how such a multi-peak distribution 
causes ambiguity in the peaks' relative heights that may not be fully corrected by the sub-sample bins as many 
of them also exhibit this second peak and tail. The other bins have similar problems but not to the extent of the 
lowest redshift bin. In the future it may be necessary to apply a further bias calibration to the clustering redshifts 
such as the Newman iteration \citep{newman08}. Another option is to apply a self-calibration technique to the 
clustering redshifts. This could take the form of constraining the relative bias by applying a corrective function to 
the summed sub-bins and the raw bins. The true galaxy bias should be a function that corrects both the sum of 
the sub-bins and the raw-bins to the same value assuming the bias does not change rapidly between sub-bins. 
We can then constrain the bias by fitting a function that brings the sum of the sub-bins and the raw bins into 
agreement. This could be considered a second-order correction on the bias after using the summing technique.


\subsection{\textsc{kdTree}, single galaxy redshifts}

\begin{figure}
    \includegraphics[width=0.5\textwidth]{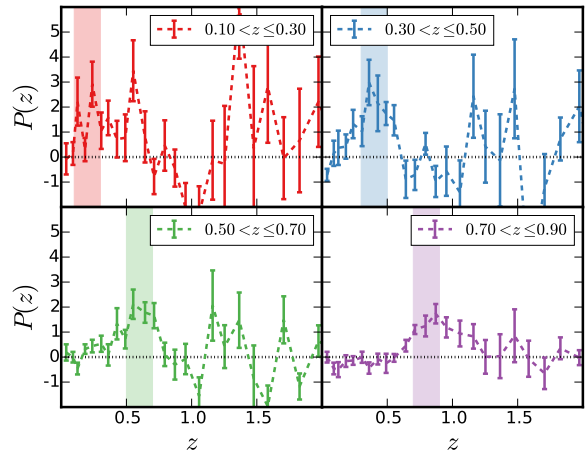}
    \caption{\label{fig:cosmos_clustering_z}
    Raw clustering-$z$s produced by \textsc{The-wiZZ} using objects from KiDS selected in $z_{\rm B}$ as the 
    unknown sample and zCOSMOS spectra as the reference sample normalized into a estimated PDF. Data 
    points colored and dashed with error bars are selections in $z_{\rm B}$ to mimic the bins of 
    \citet{hildebrandt16} (CS bins). The colored bands show the redshift selection. These clustering-$z$s show 
    that we can still use clustering redshift estimation even for small footprint surveys. The differences between 
    these clustering-$z$s and the previously shown results come from the smaller sample of reference redshifts 
    with $z>1.0$ and galaxy bias in the reference sample. It should be said that these results can not be directly 
    compared to the clustering-$z$s shown in \citet{hildebrandt16} as different scales were used and the data 
    were corrected for galaxy bias using the Newman iterative method \citep{newman08} in that study.}
\end{figure}

The method of mapping color-magnitude relations using the hybrid machine learning-clustering-$z$ method 
shows definite promise. Without adding redshift information explicitly to the \textsc{kdTree}, we were able to use 
only color-magnitude information to confirm the photo-$z$s. The peak of the single galaxy clustering-$z$s 
follows that of the photo-$z$ extremely well and with a width of the clustering-$z$ peak being no worse than
$\Delta z=0.1$ for the majority of the objects. For context, \citet{hildebrandt16} show that their 
calibrated photo-zs for similar samples to have an error of $\Delta z=0.05(1+z)$. The single galaxy clustering-
$z$s also identify some objects that have significantly narrower distributions. These single galaxy clustering-$z$s 
created with \textsc{The-wiZZ} can not only be used as a stand alone redshift estimate, but also to identify 
objects in color-magnitude space where their photo-$z$s are expected to be more reliable.

The ability to measure clustering-$z$s for individual galaxies is a powerful tool for astronomers interested in 
properties of individual or small samples of galaxies rather than statistical cosmology. We have shown how one 
can use the measured photometric properties of galaxies to select a sample of self-similar objects to estimate a 
single object clustering-$z$. Clustering-$z$s of this kind can useful for galaxies where it is difficult to estimate 
photo-$z$s  such as those without a representative training set (this applies to all photo-$z$ methods) or those 
for which galaxy spectral templates are not well understood or unavailable. This will be very common in future 
deep surveys such as LSST and WFIRST where faint galaxies will still likely be unable to have their redshift 
confirmed spectroscopically even by future 30-meter class telescopes. These kinds of single galaxy
clustering-$z$s can be used as training information for photo-$z$s for very faint objects with no redshift 
information in place of spectroscopic redshift. Samples with no optical counterpart, such as sub-millimeter 
galaxies (SMGs), also fall into this category and can be identified in redshift without the use of representative 
spectra. \textsc{The-wiZZ} can use any catalog parameter that correlates with galaxy type or redshift such as 
morphology, concentration, etc. not just flux and color making it a very general tool.


\subsection{Discussion on area for clustering-$z$s}

Clustering redshifts are largely considered a tool for current and future large area surveys. However, 
what is important for signal to noise in clustering-$z$s is not area but the product of number of reference objects 
and density of unknown objects. As such they can be used on very small area surveys with the caveat that one 
should be mindful of using very small scales and possible sample variance due to the small survey footprint. In 
Figure \ref{fig:cosmos_clustering_z} we show cluster-$z$s for the CS bins using only $0.8~\rm{deg}^2$ from 
from the intersection of KiDS and The Cosmic Evolution Survey \citep[COSMOS,][]{scoville07} using a
non-public zCOSMOS \citep{lilly09} catalog kindly provided to the KiDS team for photo-$z$ verification. The 
clustering-$z$s still show significant signal especially the higher CS bins despite the small area. For the
$z_{\rm B}>0.3$ bins, we clearly detect the redshift peaks and similar tails to that of the previous results. We 
use the same $R=0.1-1.0\, \rm{Mpc}$ radius as in the previous results. The effect of the bias of the 
spectroscopic sample can be clearly seen in the larger amplitude at $z>1.0$ compared to the results from 
SDSS and GAMA.

As an extreme case we point the reader to \citet{schrabback16} where \textsc{The-wiZZ} was used to create 
clustering-$z$s from 3D-Hubble Space Telescope survey (3D-HST) data. This data covered only $0.16~
\rm{deg}^2$ but still gave robust results thanks to the density of GRISM spectra and objects. This feature of 
clustering-$z$s will be very useful in mapping color-redshift relations out to high redshift using dense spectral 
and photometric fields with many filters such as COSMOS. Efforts to map color-redshift space such as 
\citet{masters15} which used Self-Organizing Maps (SOMs) to map a Euclid survey like color-redshift space can 
benefit from clustering redshifts, mapping out redshift degeneracies in photo-$z$ methods where more spectral 
or filter coverage will be required. In addition to this, clustering-$z$s can be used in such high redshift pencil 
beam surveys to estimate the redshift distributions of non-optically detected objects such as sub-millimeter 
galaxies.


\section{Conclusions}\label{sec:conclusions}

In this work we have presented \textsc{The-wiZZ} an open source clustering redshift estimation code designed 
to add legacy value to current and future photometric and spectroscopic surveys. The software attempts to 
make using clustering redshifts as easy as photometric redshifts are by separating out the step of computing the
2-point, cross-correlation statistics required for computing a clustering redshift for a given sample from creating 
a final clustering-$z$. \textsc{The-wiZZ} is designed for ease of use by end users of current and future surveys 
and produces clustering redshifts for any subsample of objects without the intervention of a clustering redshift 
``expert''.

We have shown robust results from both preselecting objects from a catalog (in this case photo-$z$) as well as 
from a hybrid machine learning-clustering redshift method using \textsc{kdTree}s in color-magnitude space. The 
results from the photo-$z$ selection reinforce other work that showed how preselecting objects in narrow 
redshift regions helps mitigate the effect of galaxy bias in clustering redshifts \citep{menard13, schmidt13, 
rahman16b, scottez16}. The \textsc{kdTree} clustering redshift method also shows robust results for estimating 
the redshift of individual galaxies. Such clustering redshifts are very interesting for survey users studying 
individual or small samples of objects and could possibly be used as priors for future photometric redshift codes. 
Assuming that one can measure narrow-peaked redshift distributions for a sample of individual objects one could 
use this sample as a training set for photo-$z$s. This will be especially useful for high redshift, faint objects that 
will likely not have an observable spectra even on future 30-meter class telescopes.

\textsc{The-wiZZ} will be an extremely useful clustering redshift code for future photometric surveys such as 
LSST, Euclid, and WFIRST given its speed and flexibility. These surveys are planning to rely at least in part on 
clustering redshifts to reach the precision required of their redshift distributions \citep{newman15} and a public 
code such as \textsc{The-wiZZ} can fit perfectly into these surveys collaborative software development 
environments. Future spectroscopic efforts such as the Dark Energy Spectroscopic Instrument
\footnote{\url{http://desi.lbl.gov/}} (DESI), Prime Focus Spectrograph\footnote{\url{http://sumire.ipmujp/en/}} 
(PFS), and the 4-metre Multi-Object Spectroscopic Telescope\footnote{\url{http://www.4most.eu/}} (4MOST) will 
soon provide the reference catalogs needed for this goal.

The code will likely require an amount of optimization in both the \textsc{pair\_maker} and \textsc{pdf\_maker} 
modules to minimize the size of the data file and reduce the run time per-core further. However given the 
relatively simple nature of the algorithm we have no doubt such optimizations will be found. We will continue to 
develop \textsc{The-wiZZ} over the years before these surveys begin taking data to maximize its impact. Future 
development may also include a web based tool to allow users to request the distributions for a given sample or 
incorporating \textsc{The-wiZZ} into \textsc{SQL} catalog requests. These tools will be extremely useful and will 
speed up the adoption of clustering redshifts within the community.

\textsc{The-wiZZ} will continue to be developed on
\textsc{GitHub}\footnote{\url{http://github.com/morriscb/The-wiZZ/}}. If for some reason \textsc{GitHub} closes or 
\textsc{The-wiZZ} is moved to a new repository contact the authors\footnote{E-mail: 
morrison.chrisb@gmail.com} for the location of the current repository.


\section*{Acknowledgements}

CBM would like to thank the members of the KiDS weak lensing group who helped debug the code and those 
that hosted the KiDS WL Busy Week at the Lorentz Centre in Leiden where much of this software was written: 
Catherine Heymans, Henk Hoekstra, Konrad Kuijken, and Massimo Viola. We thank the International Space 
Science Institute (ISSI) for hosting CBM and HH during which time some of the software development took 
place. We thank Mubdi Rahman for providing us with his STOMP Map of SDSS. We thank Alexander Karim, 
Benjamin Magnelli, and Jeffrey Newman for their helpful discussions. We also thank the zCOSMOS team for 
making their full, non-public redshift catalogue available to us. The authors thank the anonymous 
referee for their comments.

CBM and HH are supported by the DFG Emmy Noether grant Hi 1495/2-1. SJS is funded by grants from the 
Association of Universities for Research in Astronomy (AURA) and the Heising-Simons Foundation. MB is 
supported by the Netherlands Organization for Scientific Research, NWO, through grant number 614.001.451, 
and by the European Research Council through FP7 grant number 279396. AC acknowledges support from 
the European Research Council under the FP7 grant number 240185. PS is supported by the Deutsche 
Forschungsgemeinschaft in the framework of the TR33 `The Dark Universe'. 

This material is based upon work supported in part by the National Science Foundation through 
Cooperative Agreement 1258333 managed by the Association of Universities for Research in Astronomy(AURA), 
and the Department of Energy under Contract No. DE-AC02-76SF00515 with the SLAC National Accelerator 
Laboratory. Additional LSST funding comes from private donations, grants to universities, and in-kind support 
from LSSTC Institutional Members.

Based on data products from observations made with ESO Telescopes at the La Silla Paranal Observatory 
under programme IDs 177.A-3016, 177.A-3017 and 177.A-3018, and on data products produced by Target/
OmegaCEN, INAF-OACN, INAF-OAPD and the KiDS production team, on behalf of the KiDS consortium.

GAMA is a joint European-Australasian project based around a spectroscopic campaign using the Anglo-
Australian Telescope. The GAMA input catalogue is based on data taken from the Sloan Digital Sky Survey 
and the UKIRT Infrared Deep Sky Survey. Complementary imaging of the GAMA regions is being obtained by 
a number of independent survey programmes including GALEX MIS, VST KiDS, VISTA VIKING, WISE,
Herschel-ATLAS, GMRT and ASKAP providing UV to radio coverage. GAMA is funded by the STFC (UK), the 
ARC (Australia), the AAO, and the participating institutions.

Funding for the SDSS and SDSS-II has been provided by the Alfred P. Sloan Foundation, the Participating 
Institutions, the National Science Foundation, the U.S. Department of Energy, the National Aeronautics and 
Space Administration, the Japanese Monbukagakusho, the Max Planck Society, and the Higher Education 
Funding Council for England. The SDSS Web Site is http://www.sdss.org/.

The SDSS is managed by the Astrophysical Research Consortium for the Participating Institutions. The 
Participating Institutions are the American Museum of Natural History, Astrophysical Institute Potsdam, 
University of Basel, University of Cambridge, Case Western Reserve University, University of Chicago, Drexel 
University, Fermilab, the Institute for Advanced Study, the Japan Participation Group, Johns Hopkins 
University, the Joint Institute for Nuclear Astrophysics, the Kavli Institute for Particle Astrophysics and 
Cosmology, the Korean Scientist Group, the Chinese Academy of Sciences (LAMOST), Los Alamos National 
Laboratory, the Max-Planck-Institute for Astronomy (MPIA), the Max-Planck-Institute for Astrophysics (MPA), 
New Mexico State University, Ohio State University, University of Pittsburgh, University of Portsmouth, 
Princeton University, the United States Naval Observatory, and the University of Washington.

Funding for SDSS-III has been provided by the Alfred P. Sloan Foundation, the Participating Institutions, the 
National Science Foundation, and the U.S. Department of Energy Office of Science. SDSS-III is managed by 
the Astrophysical Research Consortium for the Participating Institutions of the SDSS-III Collaboration including 
the University of Arizona, the Brazilian Participation Group, Brookhaven National Laboratory, Carnegie Mellon 
University, University of Florida, the French Participation Group, the German Participation Group, Harvard 
University, the Instituto de Astrofisica de Canarias, the Michigan State/Notre Dame/JINA Participation Group, 
Johns Hopkins University, Lawrence Berkeley National Laboratory, Max Planck Institute for Astrophysics, Max 
Planck Institute for Extraterrestrial Physics, New Mexico State University, New York University, Ohio State 
University, Pennsylvania State University, University of Portsmouth, Princeton University, the Spanish 
Participation Group, University of Tokyo, University of Utah, Vanderbilt University, University of Virginia, 
University of Washington, and Yale University.
\\\\
{\footnotesize \textit{Author Contributions}\textrm{: The authorship list is given in three groups: the lead 
authors(CBM), followed by two alphabetical groups. The first alphabetical group includes those who are key 
contributors to both the scientific analysis and the data products. The second group covers those who have 
either made a significant contribution to the data products, or to the scientific analysis.}}


\bibliographystyle{mnras}

\bibliography{The-wiZZ}

\bsp	
\label{lastpage}
\end{document}